# Optical Network Design For 4G Long Term Evolution Distribution Network In Sleman

Firdaus, Rifa Atul Izza Asyari
Department of Electrical Engineering
Universitas Islam Indonesia
Yogyakarta, Indonesia
firdaus@uii.ac.id

Eka Indarto
Department of Electrical Engineering
Universitas Islam Indonesia
Yogyakarta, Indonesia

*Abstract*—The number of mobile users is increasing rapidly, 3GPP initiated a new technology 4G Long Term Evolution (LTE). LTE is an enabling technology as a solution to the problem of network capacity and quality in areas that have a high demand like Sleman. The quality of LTE networks are supported by backbone and distribution network. This paper describes the designing of fiber optic network as the LTE backbone network in Sleman. The backbone network requires G-652 optical cable along 85 km with a ring topology and WDM-STM64 technology. The distribution network uses GPON technology and the type of G984 optical cable along 61.35 km. The minimum of power received at the end-point in the *optisystem* simulation is -25 dBm and -26 dBm through the calculation. This value is acceptable, since it is above the minimum of power received (Receiver Sensitivity) -28 dBm which refers to the standard parameters of G.984.2. The average value of rise time is 69 ps, this value is still below the maximum allowable value of rise time 70 ps. Average BER of backbone link is $5\times10^{-4}$.

*Keywords—network design; 4G LTE, backbone, optisystem, GPON, WDM*

## I. Introduction

The public need of communication is not only in the form of voice, but also in form of video and data. It takes a communication device and network that is able to serve all the services with high quality of service. The number of smartphone users also increasing rapidly, which is predicted to reach 6.1 billion users by 2020 [1] and with the advantages of large capacity. Fiber optics became the answer to the needs of data transmission in large volumes of up to 64 Tbps and it can carried varied services in real time without buffer. Fiber optic is promising to implement as LTE backbone and distribution netwoks. Mobile operators absolute adopt this network, especially for the backbone network.

Nafiz et al in 2012 designed LTE radio network planning in Dhaka City. The step involves choosing propagation model, determine threshold of link budget, create a detailed plan based on a threshold radio, check the capacity of the network against estimated traffic, configuration planning, site survey, and planning eNodeB parameters [2]. Another study conducted by the Sri Ariyanti, which determined the frequencies used for LTE 4G technology in Jabodetabek. This study aims to give a picture of the site required for the application of LTE technology at a frequency of 1800 MHz and 2100 MHz [3]. Then the research conducted by Zakaria in Tripoly, radio network design through several stages: site survey, planning the frequency, the link budget, coverage and capacity planning. The area is divided into three areas: dense urban, urban and suburban. The frequency used is 1800 MHz with a bandwidth of 20 MHz. It use LTE FDD and a soft frequency reuse (SFR 1*3*1) with normal cyclic prefix. The study also measured the performance of LTE FDD for uplink and downlink by comparing the modulation of QPSK, 16QAM and 64QAM. Based on the simulation, it can concluded that the BER vs SNR and BLER vs SNR varies depending on several parameters such as modulation scheme, code rate and antenna configuration [4]. Latest research on the planning of 4G LTE is done by Fahri Rian, the optimization of 4G radio networks using genetic algorithms in Sleman. He use genetic algorithms as an alternative to seeking a optimal combination of existing BTS to design 4G LTE network in Sleman [5]

This paper explains the design process of optical networks as distribution and backbone networks of LTE in Sleman, which one of the biggest city in Yogyakarta, Indonesia. Sleman needs LTE network because it has a lot of population and universities.

## II. System Design

The design process of this system can be seen in figure 1. The first step is a field survey and calculation of required traffic. The second step is to design a network topology that will be used. The next step is the selection of light sources and detectors as well as the wavelength of light. Then do the calculation and simulation of power link budget and rise time budget. Last step is to evaluate whether the values are in accordance with the standards set.

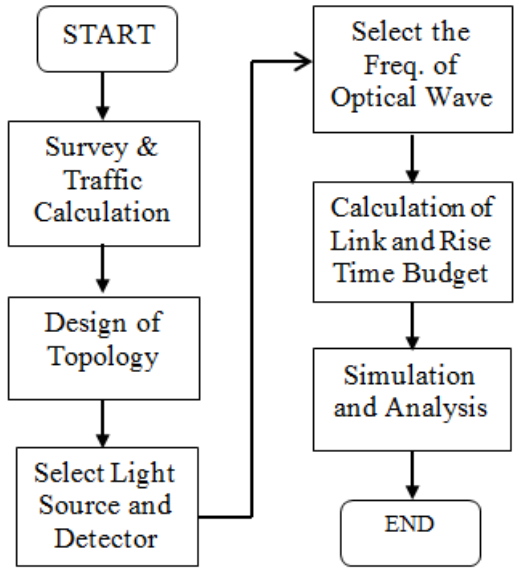

Figure 1. Design process of the system

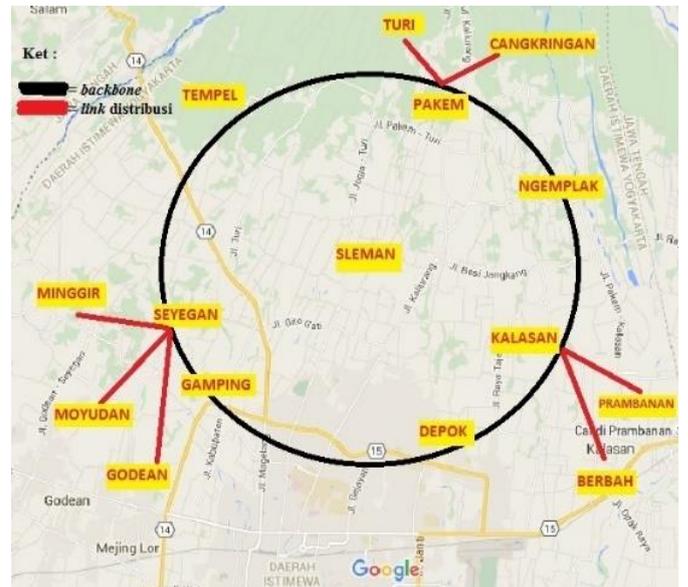

Figure 2. Proposed transmission and backbone network

## A. Traffic Calculation

Traffic calculation is done based on literature of Sleman statistic. This calculation needs the number of population and cellular penetration in Sleman. The detail of data and calculation can be seen at table 1. Traffic value will be the basis for network design. Assuming the customer growth of operator A is 5.1%. Then 5 years later, the total number of mobile users become 137378.

TABLE I. TRAFFIC PREDICTION OF SLEMAN

| No | Variabel | Value | Symbol & Formula |
|---|---|---|---|
| 1 | Sleman Population [6] | 850221 | A |
| 2 | Cellular penetration | 150% | B |
| 3 | Estimation the number of mobile subscribers | 1275331 | C=A x B |
| 4 | Penetration of operator A [7] | 42% | D |
| 5 | Customers estimation of operator A | 535639 | E=C x D |
| 6 | LTE service penetration | 20% | F |
| 7 | LTE subscribers assumptions of operator A | 107128 | F x E |

## B. Design of Optical Network Route

The design of optical networks for the distribution network of 4G LTE need coverages and routes of network to calculate power budget. Figure 2 below is the proposed optical backbone and transmission network area in Sleman.

Planning parameters used in this network, customized to the applicable standards in the ITU-T G 6:55. Planning parameters of optical fiber can be seen in Table 2.

## C. Link Budget and Rise Time Planning

Detail of link power budget describe in figure 3.

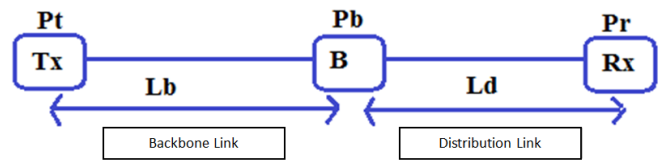

Figure 3. Illustration of link budget on backbone and distribution network

The minimum power (receiver sensitivity) on the downlink receiver (Pr) is -21 dBm (refer to the standard parameters of ITU-T G.984.2). Then based on calculation, maximum loss on distribution link (Ld) is 16.67 dB, so the minimum power on backbone ($Pb_{min}$) = Pr + Ld = -21 + 16.67 = -4.33 dBm. Transmit power on backbone (Pt) is 9 dBm, so maximum loss on backbone link (Lb) = Pb – Pt = -4,33 – 9 = -13,33 dB.

Loss on backbone link (based on network planning) is calculated using formula 1.

$$Lb = (\alpha_c \times Nc) + (\alpha_f \times L) + (\alpha_s \times Ns) + Ms \quad (1)$$

$= (0.3\times14)+(0.3\times84.9)+(0.05\times46)+3$

$= (4.2)+(25.47)+(2.3)+3$

$= 34.97 \text{ dB}$

The real loss on backbone link is greater than the maximum loss, so it required power amplifier = real loss – maximum loss = 34.97 – 13.33 = 21.64 dB. Each EDFA can give gain up to 20 dB, we need 2 pieces of EDFA to againt the loss of 21.64 dB. So the power on Rx can be calculated using formula 2.

TABEL 2. DESIGN PARAMETER OF OPTICAL NETWORK

| Parameters | Value |
|---|---|
| *Bit rate* (B) | 10 Gbps (STM-64) |
| Distance of *link* (L *link*) | 84,9 km |
| Modulation format | NRZ |
| Operation signal wave | 1550 nm |
| Margin (Ms) | 3dB |
| Total connector (Nc) | 14 |
| *Single Mode* : ITU-T G6.55 | |
| Attenuation ($\alpha_f$) | 0,3 dB/km |
| Dispersi kromatik (D) | 3,5 ps/nm.km |
| **Transmitter** | |
| *Rise time* ($t_{tx}$) | 60 ps |
| Lebar spectral ($\alpha_\lambda$) | 0,1 nm |
| Power transmit ($P_{tx}$) | 9 dBm |
| *Receiver* | |
| Rise time ($t_{rx}$) | 35 ps |
| Minimum sensitivity ($P_{rx}$) | -38 dBm |
| **Additional Components** | |
| Connector Attenuation($\alpha c$) | 0,3 dB/konektor |
| *Splice Attenuation*($\alpha s$) | 0,05 dB/*splice* |
| *Gain EDFA* (G) | 20 dBm |

$$Pr = Pt - Loss\ total + Gain\ amplifier \quad (2)$$
$$= 9 - 34.97 - 16.67 + 40$$
$$= -2.64 dBm$$

So the result of the design of the power is in accordance with the standards specified, dB stands for decibels, a unit level comparison signal, if the value is positive then called the amplification factor (gain), if the value is negative is called damping (loss). The unit dBm is the power level used.

Maximum risetime is 70% of the bit period NRZ (non-return-to-zero), it can be calculated use formula 3. BR is bit rate (10 Gbps), so the maximum risetime is 70 ps. Then we can calculate the rise time system based on formula 4, formula 5, and table 2. Number of splice (Ns) can be found by using formula 6, with $L_{cable}$ is 3 km. The results of rise time and number of splices can be seen at table 3.

$$t_s = 0.7 * (1/BR) \quad (3)$$
$$t_f = D.\sigma_\lambda.L \quad (4)$$
$$t_r = (t_{tx}^2 + t_{rx}^2 + t_f^2)^{1/2} \quad (5)$$
$$Ns = (L_{link}/L_{cable}) + 2 \quad (6)$$

TABLE 3 RISE TIME AND NUMBER OF SPLICE

| Link | Rise time (ps) | Number of Splice |
|---|---|---|
| Seyegan -Tempel | 69,552 | 6 |
| Tempel-Pakem | 69,773 | 9 |
| Pakem-Ngemplak | 69,541 | 6 |
| Ngemplak-Kalasan | 69,524 | 5 |
| Kalasan-Depok | 69,606 | 7 |
| Depok-Gamping | 69,625 | 7 |
| Gamping-Seyegan | 69,582 | 6 |

From the calculation of rise time budget, the result for the overall rise time for Link Sleman meet the rise time value system. The maximum value of rise time that the reference system is 70 ps and the above calculation is feasible to implement.

III. RESULT AND DISCUSSION

Simulations using software Optisystem-10, first performed to determine the backbone link loss then the next stage simulated distribution links based on parameters in table 4. Then will be known whether the transmitted signal is appropriate or not and to know the loss on the backbone link that has been designed. Table 4 below are the parameters of fiber optic cable.

TABEL 4 PARAMETER FIBER OPTIC CABLE

| Description | Value |
|---|---|
| Wavelength | 1550 nm |
| Distance | 85 km |
| Loss | 0,35 dB/km |
| Dispertion Type | constan |
| Dispertion | 1,8-6 ps/nm/km |
| Slope Dispertion | 0,075 ps/nm/km |
| Slot Differential | 0,2 Ps/km |
| Model type | *scalar* |
| Propagasi Type | *exponential* |
| Calculation Type | *noniterative* |
| None iteration | 2 |
| Step | *variabel* |
| Condition | *scalable* |
| Filter | 0,05 |
| Low Calculation | 1200 nm |
| High Calculation | 1700 nm |

After the data obtained and further we incorporated into the simulation Optisystem. Figure 3 below is a simulation with Optisystem to determine the link loss on the network backbone.

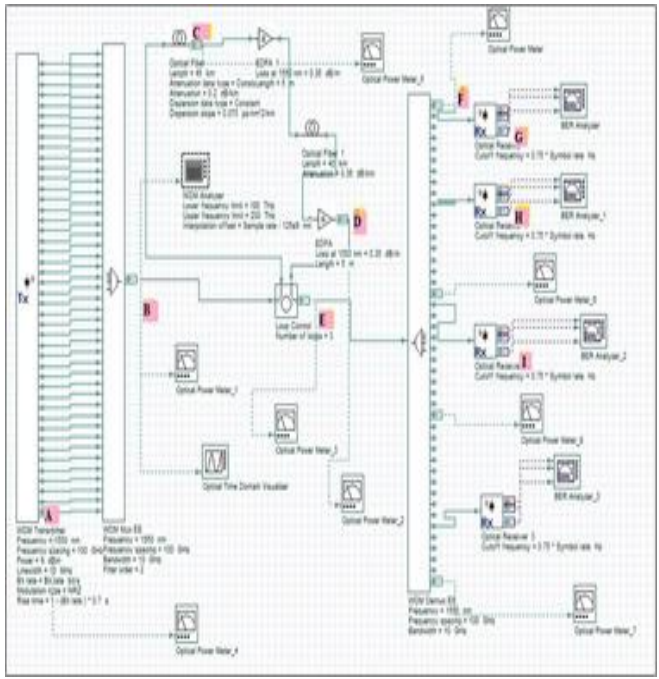

Figure 4 *Backbone link loss* simulation with *optisystem*

The total power on Optisystem simulations prove that the system has been designed properly and in accordance with the parameters have also been the result of output design. Output power link backbone at node A is 11.436 dBm, at node B is -14.367 dBm, at node C is -42.212 dBm, and at node D is -52.838 dBm. We also evaluate value of bit error rate (BER) at end point (node G, H, I at figure 4), they are $5 \times 10^{-4}$, $6 \times 10^{-3}$, and $6 \times 10^{-6}$. After evaluated backbone link, then we try to evaluate distribution link using optisystem simulation, the parameter system can be seen at table 5.

TABLE 5. PARAMETER FOR DISTRIBUTION LINK SIMULATION

| Device | Nilai |
|---|---|
| Fiber optik | Loss = 0,2 dB/km |
|  | Dispertion = 16,75 ps/nm/km |
|  | Distance = 2-20 km |
| Laser diode | Power = 10 dBm |
| Photodetector | Responsivity = 0,9 A/W |
| Band-pass filter | 0,75*bit rate |
| Splitter bidirectional | 1x2 – 1x4 |

## IV. CONCLUSIONS

The design of link distribution network 4G LTE in Sleman using GPON technology need type of G984 optical cable along the 61.35 km. The received power at the end-point (ONU) in the simulated optisystem is -25.010 dBm and it is -26.626 dBm through the real calculation result. This value is still good performance, because it is smaller than minimum value of power (Receiver Sensitivity) -28 dBm which refers to the standard parameters of ITU-T G .984.2 for GPON distribution links. Average of rise time is 69 ps, it is still under the maximum rise time 70 ps. BER of backbone links are $5 \times 10^{-4}$.

ACKNOWLEDGMENT

This work has been supported by Direktorat Penelitian dan Pengabdian Masyarakat (DPPM) and Departmen of Electrical Engineering Universitas Islam Indonesia.